# Hierarchy of Electronic Properties of Chemically Derived and Pristine Graphene Probed by Microwave Imaging

[1]Worasom Kundhikanjana, [1]Keji Lai, [2]Hailiang Wang, [2]Hongjie Dai, [1]Michael A. Kelly, [1]Zhi-xun Shen

[1] *Geballe Laboratory for Advanced Materials, Departments of Physics and Applied Physics, Stanford University, CA 94305*
[2]*Department of Chemistry, Stanford University, Stanford, CA 94305*

**Abstract** – **Local electrical imaging using microwave impedance microscope is performed on graphene in different modalities, yielding a rich hierarchy of the local conductivity. The low-conductivity graphite oxide and its derivatives show significant electronic inhomogeneity. For the conductive chemical graphene, the residual defects lead to a systematic reduction of the microwave signals. In contrast, the signals on pristine graphene agree well with a lumped-element circuit model. The local impedance information can also be used to verify the electrical contact between overlapped graphene pieces.**

The unique electronic band structure of graphene, the single layer two-dimensional form of carbon, has led to numerous fascinating discoveries[1-4]. Because of its novel electronics properties and potential for device function down to a single atomic layer, graphene has been proposed for future applications in electronics[5,6] and spintronics[7]. However, most of the remarkable properties are intimately tied to the exquisite material quality of the pristine graphene (PG) obtained by mechanical exfoliation of bulk graphite[1-3], which clearly is not a scalable method. Therefore, significant research efforts have been devoted to large-scale production of graphene through epitaxial growth[8,9] or chemical process[10-17]. For example, the hydrazine reduction process involves the oxidation and chemical exfoliation of graphite into single-layer graphite oxide (GO) in solution, and the subsequent thermal (annealed GO or AGO) or chemical (reduced GO or RGO) treatment to remove excess functional groups[10-12,17]. Another process directly inserts molecules into the intercalated graphite to obtain graphene sheets (GS), and then anneals (AGS) them for better quality[13,14]. In the microscopic scale, the level of impurities in these materials varies drastically, resulting in a broad range of conductivity (σ) and a hierarchy of the electrical properties.

In order to investigate the disorder effects in the graphene films, local electrical imaging with scanning probe microscopy is indispensable. The electronic inhomogeneity in graphene has been studied by a number of DC or low-frequency AC probes[18-22]. For high-speed device applications, on the other hand, the local microwave response is also important. In this case, a near-field scanning probe is imperative to obtain sub-micron spatial resolution. Besides, unlike DC probes, contact electrodes are not required for microwave imaging, allowing fast and convenient measurements on a large amount of specimens for statistic analysis.

The schematic of our microwave impedance microscope (MIM)[23,24] setup is shown in Fig. 1a. The probe is a modified atomic-force microscope (AFM) cantilever with Al transmission lines patterned on the $Si_3N_4$ base. The state-of-the-art AFM platform (Park AFM, XE-70) allows a small contact force (~ 2nN) and a low topographic noise (< 1nm) in the contact mode operation. The nominal spatial resolution ~100nm is limited by the diameter of the focused ion beam deposited Pt tip. The two 90° out-of-phase channels from the microwave electronics, MIM-C and MIM-R, record the small variations of the imaginary (capacitive) and real (resistive) components of the effective tip impedance $Z_{tip}$, as shown in Fig. 1b. The color rendering in this paper is such that higher tip-sample capacitance (loss) corresponds to brighter color in the MIM-C (MIM-R) images. The chemically derived GO/AGO/RGO and GS/AGS samples in this study are deposited on 100nm $SiO_2$/p++Si and PG on 300nm $SiO_2$/p++Si substrates.

Typical MIM images of the pristine graphene with patterned gold electrodes are shown in Fig. 1c, as well as the simultaneously taken topographic image in the inset of Fig. 1a. While the thick electrodes (~100nm) completely overshadow the thin PG pieces (<1nm) in the AFM image, strong microwave signal on the graphene, same as that on the gold electrode, is observed in the MIM-C image. The marked difference between Fig. 1a and 1c is a direct manifestation of the capability of the microwave probe to perform pure electrical imaging. In fact, as detailed later, the PG pieces and the electrodes are electrically shorted together so they exhibit the same color in MIM-C. The MIM-R image, on the other hand, displays almost no contrast over the substrate, except at the edges of the electrodes where tunneling resistance between the tip and large metal pieces induces finite loss.

Since the materials studied here cover a conductivity range over many orders of magnitude, it is essential to first obtain a unified picture of the MIM responses [25]. In Fig. 2a, the MIM-C and MIM-R signals on a thin sheet are plotted as a function of $\sigma$.[26] For extremely resistive ($\sigma < 10^0$ S/m) pieces, very weak microwave signals are expected. In fact, the MIM-C channel will show slightly negative contrast over the substrate because the film reduces the tip-sample capacitance by effectively increasing the distance from the tip to the metallic p++Si. As $\sigma$ increases, both MIM-C and MIM-R signals increase due to the reduction of the sheet resistance $R_S$. At our working frequency 1GHz, $R_S$ becomes comparable to the sample-substrate capacitance at around $\sigma_c \sim 10^3$ S/m, where the loss channel MIM-R signal reaches a maximum. For higher $\sigma$, the resistive loss in the film decreases, resulting in a decreasing MIM-R response. The MIM-C signal, on the other hand, keeps increasing till $\sigma \sim 10^5$ S/m, beyond which the microwave signal stays constant. At this conductive or lossless limit, a nano-sheet on the SiO$_2$/p++Si substrate is equivalent to a parallel-plate capacitor. The MIM-C signal is then proportional to the area of the piece.

As we overlay the reported DC conductivity[10, 12-14, 17] of all species in this study on top of the simulated curves in Fig. 2a, a hierarchy of the microwave response is readily expected. Graphite oxide[10] and the derivatives AGO[10,13] and RGO[12,14,17] are found on the left side of the spectrum with low conductivities. In Fig. 2b, high-density GO clusters (>10nm roughness, topography not shown) appear darker than the substrate in MIM-C as discussed above. The MIM-R images are featureless as expected. After annealing, the DC conductivity of AGO is considerably higher than that of the GO. As seen in Fig. 2c, while most AGO pieces still appear dark in the MIM images, certain regions of the samples do show discernible positive contrast over the substrate, corresponding to $\sigma = 10^1 \sim 10^2$ S/m. For the slightly better quality RGO samples (Fig.

2d) with σ = $10^2$ ~ $10^3$ S/m, the MIM images in both channels are brighter over the background. We note that both AGO and RGO are strongly inhomogeneous in the local conductivity, which is vividly shown by overlaying the actual dimension of the pieces measured from the topography (dotted lines in Fig. 2c and 2d) to the MIM images.

The GS/AGS/PG samples occupy the high-σ side in Fig. 2a. For GS/AGS pieces with conductivity in the order of $10^4$ ~ $10^5$ S/m,[13,14] the MIM-C signal should mostly saturate at the conducting limit, while appreciable loss is present in the MIM-R channel. In general, the simulation result fits well to the images in Fig. 2e and 2f, and we do not observe obvious difference between these two species. Interestingly, certain internal features are clearly identified, such as some dark spots in the MIM-C images and bright fringes in the MIM-R channel. This observation is in sharp contrast to the homogeneous images obtained on the PG samples (σ > $10^6$ S/m) in Fig. 2g, which by far exhibit the best quality.

Further insights on GS/AGS samples can be obtained by statistic analysis on a large number of specimens. In Fig. 3a and 3c, we plot the size dependence of the average MIM-C signals of many GS/AGS and PG sheets. The area of each flake is extracted by the threshold grain detection. The error bars represent the actual signal fluctuations in the samples rather than the instrumental noise. Using the finite-element analysis (FEA) software COMSOL 3.5,[24,25] we directly calculate the tip-sample impedance (Fig. 3b) as a function of the area of the graphene pieces. The results, using the nominal conductivity ($10^5$ S/m for GS/AGS and $10^7$ S/m for PG), are included in Fig. 3a and 3c for comparison. While the simple lumped element circuit model accurately describes the microwave response on high quality PG sheets, the measured MIM-C data of GS/AGS samples systematically fall below the theoretical curve. And the deviation is too large to be accounted for by any inaccuracy in the simulation parameters, e.g., the $SiO_2$ layer thickness. We ascribe such effect to the intrinsic nonuniformity in these materials. For instance, electronic defects from the chemical process usually result in low σ near the edges[13]. Inferior local conductivity or a residual insulating overlayer could also drastically reduce the MIM-C signal (dotted line in Fig. 3a). Note that the measured data cannot be reproduced by assuming a low ($\sigma < 10^4$ S/m) but homogeneous "apparent conductivity". In conclusion, the discrepancy in Fig. 3a is a direct consequence of the local inhomogeneity in the chemically derived graphene.

Careful examination on the GS/AGS data also reveals some interesting features in these materials. One of the potential applications for the chemically derived graphene sheets is to form conductive thin films[12,15,16]. For this purpose, characterization of good electrical connections across partially overlapped pieces is crucial. If two high-$\sigma$ pieces in close proximity are indeed shorted together with very low contact resistance, they would appear as the same capacitive impedance in parallel to $Z_{tip}$ and display the same color in the MIM-C image. While most overlapped pieces are well joined electrically, exceptions are found. For example, the GS pieces showing high MIM-C signals (bright) in both Fig. 4a and 4b are apparently isolated from the dark ones right next to them by large impedance, which is hard to identify by looking only at the topographic images. As discussed in Fig. 3, the MIM-C images signal of electrically joined GS pieces is now determined by the total area rather than that of the individual flakes. Again, such electrical information of large-area conductive graphene sheets is critical but extremely difficult to obtain by other means.

**Acknowledgements**

We thank K. Todd for the assistance with pristine graphene, D. Goldhaber-Gordon for useful discussions, and C. Buenviaje-Conggins for the instrumental advice. The research is supported by Center of Probing the Nanoscale (CPN), Stanford University, gift grants from Agilent Technologies, Inc., and DOE contract DE-FG03-01ER45929-A001. This publication is also based on work supported by Award No. KUS-F1-033-02, made by King Abdullah University of Science and Technology (KAUST) under the global research partnership (GRP) program. CPN is an NSF NSEC, NSF Grant No. PHY-0425897. The work on graphene synthesis is supported by MARCO-MSD, Intel and ONR.

26. Strictly speaking, since the thickness $t$ of the graphene samples is much smaller than the tip diameter, it is the sheet resistance $R_S = 1/(\sigma \cdot t)$ that characterizes the microwave response. For a better description and comparison with the literature, we assume $t \sim 1$nm for the samples and convert $R_S$ to $\sigma$ accordingly throughout the paper.

**Figures:**

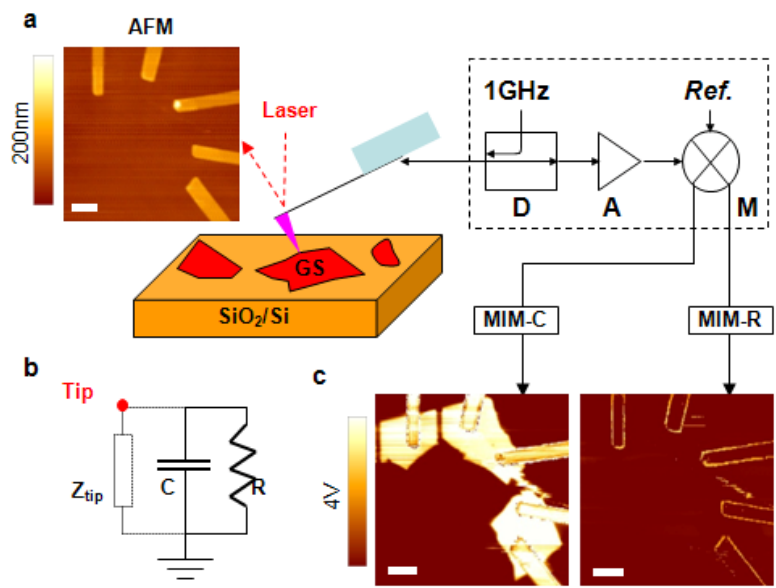

**Figure 1 Microwave impedance microscopy setup and typical microwave images on graphene. a.** Schematic of the AFM-compatible MIM setup. The reflected 1GHz signal from the tip is amplified and demodulated by the microwave electronics in the dashed box (D – directional coupler, A – amplifier, M – mixer). Inset – AFM topography of several peel-off graphene pieces contacted by gold electrodes. **b.** Equivalent lumped-element circuit, showing the tip ($Z_{tip}$) and sample impedance (R and C). **c.** MIM images of PG sheets with electrodes, showing striking difference between the AFM and MIM signals. The RF gain for this set of images is lower than that used for the rest of the paper to avoid saturation of the output. Scale bars in **a** and **c** are 2 μm.

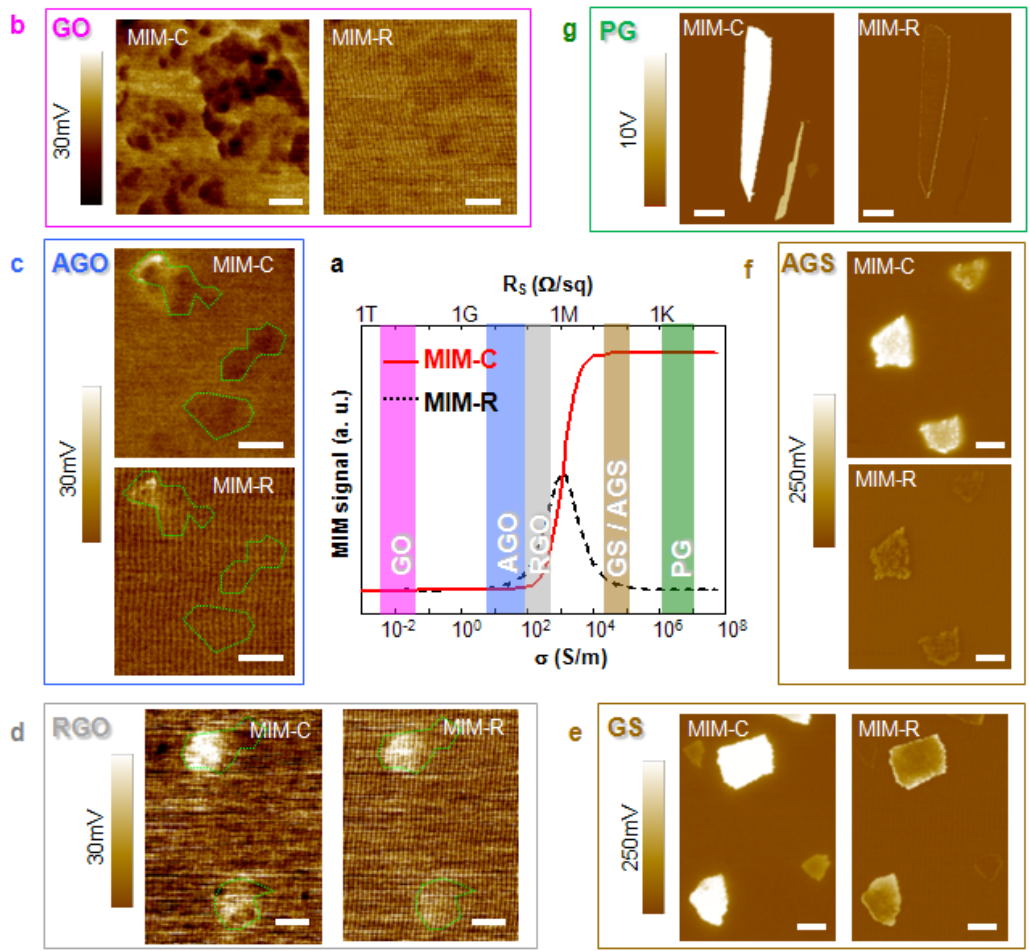

**Figure 2 Hierarchy of local conductivity and microwave images of chemically derived and pristine graphene. a.** Simulated MIM response as a function of the conductivity of a thin sheet of specimen, 1nm in thickness and 1μm in diameter, placed on a 100nm $SiO_2$/p++Si substrate. The reported DC conductivity range of all species in this study is highlighted in different colors. **b-g.** Counterclockwise from the top left corner, MIM images of GO, AGO, RGO, GS, AGS, and PG samples, respectively. Refer to the text for detailed descriptions. The dotted lines in **c** and **d** encircle the actual size of the pieces measured from the topography (see Supp. Info. S3 for the AFM images). All scale bars are 200nm except for 1 μm in **g**.

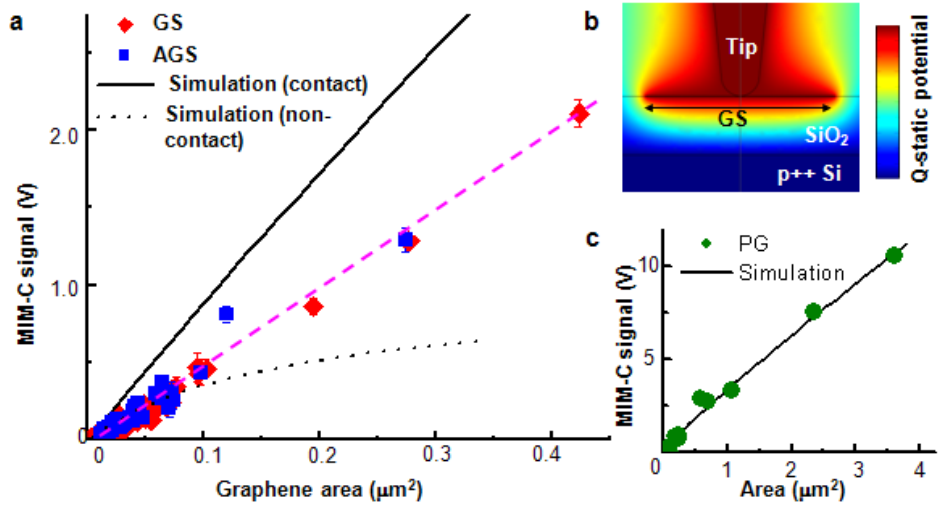

**Figure 3 Statistic analyses of the microwave signals and comparison with modeling. a**. Size dependence of the MIM-C signals on many GS (red) and AGS (blue) pieces. The dashed line is a guide for the eyes. The solid line shows the simulation results when the tip is in contact with conductive sheets ($\sigma = 10^5$ S/m). The dotted line corresponds to the simulation results where the tip is separated from the graphene by 0.5nm. The measured data lie between these two cases. **b.** FEA simulation of the quasi-static potential distribution when the tip is in contact with a conductive graphene sheet on the $SiO_2$/p++Si substrate. **c**. MIM-C signals for the PG pieces, fitting nicely to the simulated curve.

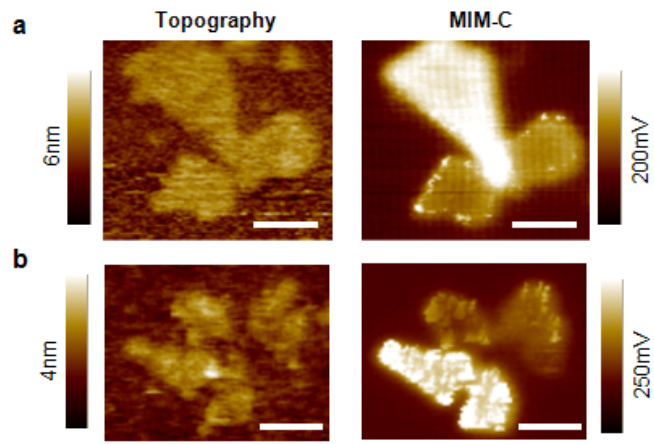

**Figure 4 Electrical connection between overlapped graphene sheets. a.** and **b.** Examples of topographic and MIM images of GS. While it is difficult to identify from the former whether overlapped graphene sheets are shorted together, the MIM-C images clearly display the electrically connected pieces as the same color. All scale bars are 200nm.